\documentclass{aa}  
\usepackage{graphicx}
\usepackage{mathtools}
\usepackage{caption}
\usepackage{subcaption}
\usepackage{natbib}
\bibpunct{(}{)}{;}{a}{}{,} 
\usepackage{txfonts}
\begin{document}

   \title{LOFAR Observations of Swift J1644+57 and Implications for Short-Duration Transients}


\author{Y.~Cendes\inst{1}
\and R.A.M.J.~Wijers\inst{1}
\and J.D.~Swinbank\inst{2}
\and A.~Rowlinson \inst{3}
\and A.J.~van~der~Horst\inst{1}
\and D.~Carbone\inst{1}
\and J.W.~Broderick\inst{4}\inst{5}
\and T.D.~Staley \inst{4}
\and A.J.~Stewart\inst{4}
\and G.~Molenaar\inst{1}
\and F.~Huizinga \inst{1}
\and A.~Alexov \inst{6}
\and M.E.~Bell\inst{3}
\and T.~Coenen\inst{1}
\and S.~Corbel\inst{7}\inst{8}
\and J.~Eisl\"offel\inst{9}
\and R.~Fender\inst{1}
\and J.~Grie{\ss}meier\inst{10}\inst{8}
\and J.~Hessels\inst{11}\inst{1}
\and P.~Jonker\inst{12}
\and M.~Kramer\inst{13}
\and M.~Kuniyoshi\inst{14}
\and C.J.~Law\inst{15}
\and S.~Markoff\inst{1}
\and M.~Pietka\inst{4}
\and B.~Stappers\inst{16}
\and M.~Wise\inst{11}\inst{1}
\and P.~Zarka\inst{17}
}

\offprints{Y. Cendes, \email{Y.N.Cendes@uva.nl}}

\institute{Anton Pannekoek Institute, University of Amsterdam, Science Park 904, 1098 XH Amsterdam, The Netherlands
\and Department of Astrophysical Sciences, Princeton University, Princeton, NJ 08544, USA
\and CSIRO Astronomy and Space Science, PO Box 76, Epping, NSW 1710, Australia
\and Astrophysics, Department of Physics, University of Oxford, Keble Road, Oxford OX1 3RH, UK
\and School of Physics and Astronomy, University of Southampton, Highfield, Southampton SO17 1BJ, UK
\and Space Telescope Science Institute, 3700 San Martin Dr, Baltimore, MD 21218
\and Laboratoire AIM (CEA/IRFU - CNRS/INSU - Universit\'e Paris Diderot), CEA DSM/IRFU/SAp, F-91191 Gif-sur-Yvette, France
\and Station de Radioastronomie de Nan\c{c}ay, Observatoire de Paris, CNRS/INSU, USR 704 - Univ. Orl\'eans, OSUC, 18330 Nan\c{c}ay, France
\and Th\"uringer Landessternwarte, Sternwarte 5, D-07778 Tautenburg, Germany
\and LPC2E - Universit\'{e} d'Orl\'{e}ans / CNRS
\and ASTRON, the Netherlands Institute for Radio Astronomy, Postbus 2, 7990 AA Dwingeloo, The Netherlands
\and SRON, Netherlands Institute for Space Research, Sorbonnelaan 2, 3584 CA, Utrecht, the Netherlands
\and Max-Planck-Institut f\"{u}r Radioastronomie, Auf dem H\"{u}gel 69, 53121 Bonn, Germany
\and NAOJ Chile Observatory, National Astronomical Observatory of Japan, 2-21-1 Osawa, Mitaka, Tokyo 181-8588, Japan
\and Department of Astronomy and Radio Astronomy Lab, University of California, Berkeley, CA, USA
\and Jodrell Bank Centre for Astrophysics, School of Physics and Astronomy, The University of Manchester, Manchester M13 9PL, UK
\and LESIA, Observatoire de Paris, CNRS, UPMC, Universit\'e Paris-Diderot, 5 place Jules Janssen, 92195 Meudon, France}

   \date{}

 
  \abstract
   {}
   {We have analyzed low frequency radio data of tidal disruption event (TDE) Swift J1644+57 to search for a counterpart.  We consider how brief transient signals (on the order of seconds or minutes) originating from this location would appear in our data.  We also consider how automatic radio frequency interference (RFI) flagging at radio telescope observatories might affect these and other transient observations in the future, particularly with brief transients of a few seconds duration.}
   {We observed the field in the low-frequency regime at 149 MHz with data obtained over several months with the Low Frequency Array (LOFAR).  We also present simulations where a brief transient is injected into the data in order to see how it would appear in our measurement sets, and how it would be affected by RFI flagging.  Finally, both based on simulation work and the weighted average of the observed background over the course of the individual observations, we present the possibility of brief radio transients in the data.}
   {Our observations of Swift J1644+57 yielded no detection of the source and a peak flux density at this position of 24.7 $\pm$ 8.9 mJy.  Our upper limit on the transient rate of the snapshot surface density in this field at sensitivities < 0.5 Jy is $\rho < 2.2 \times10^{-2}$ deg$^{-2}$.  We also conclude that we did not observe any brief transient signals originating specifically from the Swift J1644+57 source itself, and searches for such transients are severely limited by automatic RFI flagging algorithms which flag transients of less than 2 minutes duration.  As such, careful consideration of RFI flagging techniques must occur when searching for transient signals.}
   {}


   \maketitle
%

\section{Introduction}

A tidal disruption event (TDE) is a transient phenomenon that occurs when a star passes too close to a black hole at the center of a galaxy and is disrupted by tidal forces \citep{Rees1988}.  In a TDE, about half of the star becomes unbound from the system, while the remainder of the stellar mass is bound in highly eccentric orbits \citep{Metzger2011}.  This mass then returns to the vicinity of the black hole on a range of timescales and accretes at a rate $\dot{M}$, which at first can exceed the Eddington limit until the initial flare decreases with a power law relationship $\dot{M} \propto t^{-5/3}$ \citep{Metzger2011}.  This initial flare produces emissions visible in the X-ray, optical, and UV wavelengths \citep{Ulmer1999}.  

Relativistic outflows are often observed when black hole accretion takes place, such as onto supermassive black holes (SMBH) in active galactic nuclei \citep{Rees1988}.  However, the details for producing such a jet are not yet well understood, and it is not yet certain whether all TDEs have relativistic jets or if this only occurs in special conditions \citep{Bower2013}.  Thus, studying a TDE can provide insights into the conditions required to create a relativistic jet, and how these jet properties can change as $\dot{M}$ evolves.

Radio emission can originate from the relativistic jet in a TDE as the material in the jet interacts with the circumnuclear medium around the SMBH \citep{Metzger2011}.  The emission originates from both the forward shock as the jet interacts with the circumnuclear medium and the reverse shock which propagates back through the outflow.  The emission from these components depends on the jet's lifetime, the density of the circumnuclear medium surrounding the system, and the energy released in the electrons and the magnetic field.  The radio emission itself peaks when the outflow has slowed down to mildly relativistic speeds, on a time scale of $t \sim1$ yr.  Eventually, the blast wave from the TDE decelerates and the jet reduces to a spherical, non-relativistic expansion \citep{Metzger2013}.  When the jet becomes non-relativistic, radio emission occurs isotropically, meaning TDEs could be possibly detected at late times that were unobserved in other frequencies because the high-energy emission from the relativistic jet was pointed away from the observer initially.

So far there have been several candidate TDEs at X-ray and optical frequencies \citep{vanVelzen2012}, but only two TDEs have been associated with a transient radio counterpart, which were both first detected in X-rays by Swift: Swift J1644+57 \citep{Levan2011, Bloom2011, Zauderer2011}, and Swift J2058+05 (Cenko et al, 2012).  Swift J1644+57 was first detected on March 25, 2011, and subsequent observations have placed the event near the center of a compact galaxy at $z~\simeq~0.35$.  Bright radio fluctuations have been observed on the scale of months, and the observations have indicated signatures from relativistic outflows directed toward Earth and interactions between the jet and the circumnuclear medium \citep{Levan2011,Wiersema2012,Berger2012,Zauderer2011,Zauderer2013}.  Swift J1644+57 was also observed by \citet{Zauderer2013} to rebrighten at all frequencies on a timescale of about one month$-$ an unexpected observation if the blast wave from the jet is evolving with constant energy.  This could be due to several reasons, such as extra energy injected into the system when the forward shock catches up to the slower material  \citep{Berger2012} or a jet with angular structure \citep{Metzger2013}.  Extended radio observations of the object up to 600 days, combined with a sharp decline in X-ray observations during this period, indicate that the relativistic jet has turned off \citep{Zauderer2013}.

Radio observations of a TDE are also interesting at long radio wavelengths, as the signal is expected to peak in the low frequency domain years after the initial burst.  From their properties, it is predicted that TDEs will be the most frequent extra-galactic transients that will be found in surveys \citep{Frail2012}, making their study particularly important for the new wide-field transient surveys coming online such as LOFAR \citep{LOFAR}, WSRT/APERTIF \citep{WSRT}, MWA \citep{MWA}, ASKAP \citep{ASKAP} and MeerKAT \citep{MeerKAT}.


Low frequency radio observations are also key to understanding the process of interaction between the relativistic winds in the outflow- especially if they are highly magnetized- and the circumnuclear material.  \citet{Usov2000} outline how gamma-ray bursts (GRBs) generated through such interactions would be accompanied by very short pulses of low-frequency radio emission which could be detected up to a few tens of MHz, and how these bursts can be random in their nature due to inhomogeneities and substructure in the material they are traveling through.  These pulses at $\nu \sim30$ MHz would have a duration of $\sim 7 \times 10^{2}$~s with a flux of $< 100$ Jy \citep{Usov2000}.  Such signals have not yet been observed, either originating from GRBs or TDE events, because extended monitoring at these frequencies before now has been unfeasible in imaging surveys.  However, they could be used to determine distances and probe the interstellar medium through their dispersion measures.  Such transient models are still speculative, and no searches for them from a TDE have as yet occurred.   Further, there are questions about just how such signals would look in radio astronomical data because of their brief duration.  Also of concern is just how such astronomical signals would be processed by automatic radio frequency interference (RFI) detection algoritms, because such pulses originating from the substructure of the jet material would be short in duration and irregular in their timing, so they could be mistaken as random bursts of RFI.

In this paper we will present first our observations of Swift J1644+57 in the low frequency regime with LOFAR, as well as results from a dedicated search for brief flashes of transient signals from this specific object in the range of minutes to seconds.  We will also show simulation work in order to model just how these signals would appear in our data, and to understand how they could be affected by RFI flagging algorithms.  Finally, we will also address the topic of transient rates within our data, both in general terms and the more specific case of when they originate from the location of Swift J1644+57.

\section{Observations of Swift J1644+57}

Observations of Swift J1644+57 were made with the LOFAR telescope as an extra target observation during routine transient monitoring observations (Fender et al., in prep) using one sub-array pointing of the telescope.  Six different observations took place over various intervals in the first half of 2013, using one beam at 149 MHz with a bandwidth of 781 kHz.  On each date six measurements of 11 minutes duration each were taken observing the Swift J1644+57 field, along with calibrator sources 3C295 and Cygnus A.  From these 36 measurements, 26 were usable for analysis yielding 4 hours of data in total, as the rest suffered technical issues, with the break down of the data used visible in Table 1.

The data were pre-processed using the standard methods with LOFAR: it was flagged for radio frequency interference (RFI) using AOFlagger \citep{AOFlagger}, and the data calibration and imaging were carried out using methods described by \citet{Heald2011} and \citet{LOFAR}.  Specifically, we used a model of the source for the calibration of the calibrator sub-bands, (see \citet{Scaife2012}), and then the gain amplitudes and phases were transferred to the target field data.  For the calibration we used a model of the source, and refined the calibration further by performing a phase self-calibration on the target field as outlined by \citet{Heald2011}.  For the calibration, we used a model of 3C295 from \citet{Scaife2012}, and a model of Cygnus A from McKean et al. (priv. comm.). We refined the calibration further by performing phase-only self-calibration on the target field, as outlined by \citet{LOFAR}. For imaging, we used the AWImager \citep{Tasse2013}; a maximum projected baseline of 12 km was chosen to ensure a reliable image from the given u,v coverage per 11 min snapshot with an angular resolution of x arcsec $\times$ y arcsec (beam position angle $z^{\circ}$).

After this was done for the individual measurement sets, we added the images from the different observations together in order to get one image from all the different data runs.  This image has a total field of view of 11.35 deg$^2$ in a circular region, though a much smaller region is shown in Figure 1 in order to highlight the position of Swift J1644+57.  From these observations, the resulting image at the location of Swift J1644+57 itself was noise-like, resulting in a non-detection.  The peak flux density at the position of the TDE is 24.7 $\pm$ 8.9 mJy at 149 MHz, measured using a forced fit of a point source in PySE (Carbone et al., in prep).

\begin{table}
\centering
\begin{tabular}{lccc}
\hline\hline
Date&Number of & Number of Used & Days after\\
 & Observations & Observations & Outburst \\
\hline
2013-02-10& 6 & 6 & 688\\
2013-03-10& 6 & 6 & 716\\
2013-03-24& 6 &6 & 730\\
2013-04-21& 6 & 4 & 758\\
2013-05-19& 6 & 4 & 786\\
2013-07-13& 6& 0 & 841\\
    \end{tabular}
\caption{\label{t7}Observations of Swift J1644+57 ordered by date.  Some of the observations were not used due to high background levels and other technical defects.}
\end{table}

\begin{figure}
\includegraphics[width=0.5\textwidth]{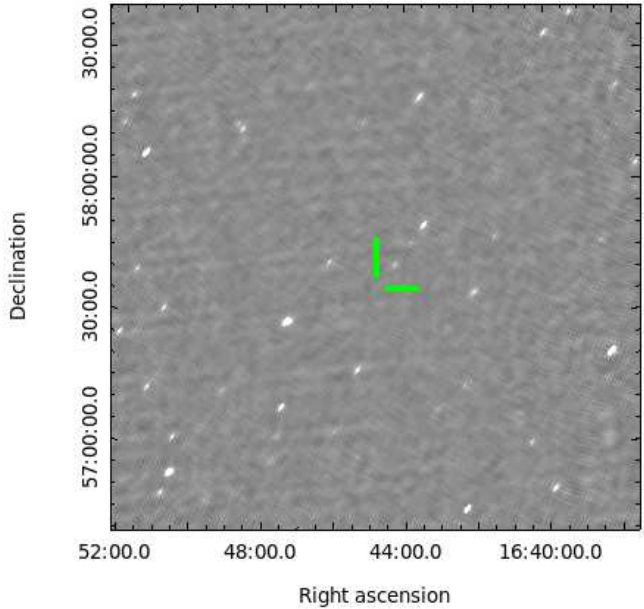}
\centering
\caption{A LOFAR 149 MHz image of Swift J1644+57 with the location of the tidal disruption event marked at center, including RA and Dec (J2000) coordinates.}
\end{figure}

\section{Short Duration Transients and RFI Flagging}

Swift J1644+57 has previously been observed in other wavelength regimes as a long-duration transient, but consideration was also given to the potential observation of shorter transient signals that could be detected with the LOFAR telescope, such as those described by \citet{Usov2000}.  One particular concern was that such signals could be mistaken for RFI during the initial processing of the data by the LOFAR Observatory.  For a modern radio telescope array such as LOFAR in a high-RFI environment, RFI flagging is an automated process due to the large amounts of data involved in order to efficiently deal with RFI.  Transient radio sources can be quick bursts of radiation that have many similarities in their signatures to what a computer algorithm may recognize as RFI and automatically flag.  An illustration of this can be seen in Figure 2, where we have the amplitudes for a data sample with a simulated transient of 30 seconds duration in it (see Section 4) visualized in the time-frequency domain.  Through the sequence described in more detail below, the AOFlagger will flag the transient automatically as contaminated data.  Thus, it was crucial to understand whether such transient signals could potentially be flagged before the data is even examined by the observer.

As briefly mentioned previously, the current default RFI flagging software for LOFAR imaging, which is used routinely during transient surveys, is called the AOFlagger.  AOFlagger is designed in order to work effectively for the most demanding projects of LOFAR in terms of noise and sensitivity, such as the LOFAR Epoch of Reionization Project \citep[EOR;][]{Jelic2014}, while trying to minimize issues with regards to accuracy and speed.  RFI typically appears as bright and sudden amplitude changes in the signal levels, and there are several steps in this automated flagging procedure to identify such signals accordingly.  At its most basic level, the AOflagger will take a sequence of amplitude information of one polarization in a single sub-band and clip out the highest amplitude information, do surface fitting of the remaining data in order to identify any fringes, clip again, and iterate over these steps again.  A more detailed description of these individual steps and why they are particularly important for transient searches are as follows:

\begin{itemize}
\item The first part of the flagger deals with the number of iterations in the entire sequence, which is necessary for finding low-level RFI. However, because these iterations can be costly in time, they are usually kept to a minimum when flagging, and tests on data from several telescopes have shown that two iterations are sufficient for an accurate fit.

\item SumThreshold is a combinatorial thresholding algorithm where the threshold $\chi_{M}$ is determined by M, the number of samples in the data set \citep{SumThreshold}. Thresholding is a system where signals whose values exceed $\chi_{M}$ are flagged as RFI, where the threshold can be defined based on the average values of a signal in a given data set.  As explained in detail by \citet{SumThreshold}, what makes SumThreshold unique is that it processes data in order of decreasing thresholds $\chi_{1}, \chi_{2}, ..., \chi_{M}$, where the first threshold is the highest and calculated from the averages of all the data.  Processing thresholds in this way is advantageous because if a sample is already flagged as RFI by an earlier threshold cutoff this contaminated data will not be included in the sum to calculate the average threshold level, whereas an algorithm processing data in a different order may mistake a strong astronomical signal in the data as interference.

SumThreshold is performed in an iteration once before the surface fitting step in order to ignore the RFI when fitting, then once again after the fitting is established in order to do the actual flags.  SumThreshold is processed in both the time and frequency domains.  As such, processing the data in both these domains can be an issue for particularly bright, brief transients.

\item Next there is a channel and time selection step which flags problematic channels and time steps which may be fully contaminated but are not yet flagged. This is done by comparing the RMS values within a channel and flagging completely any channels where the amplitude of the signal exceeds a standard deviation $>$ 3.5 compared to the surrounding amplitudes. This step is implemented in order to have convergence more quickly within the algorithm, though it may inaccurately flag data that is not RFI.  For example, a sudden, brief transient in an otherwise quiet radio field would have its amplitude suddenly change by a high value of $\sigma$ compared to the data around it, and it could be mistaken for RFI and flagged at this stage.

\item Surface fitting then occurs, which removes fringes from strong sources in order to increase accuracy. This is done with a Gaussian kernel sliding window in both time and frequency space, and is a time-consuming step when compared to the other steps listed here.  As such, these time-consuming steps can pose a problem as transients increasingly focus on real-time data analysis with instruments such as AARTFAAC \citep{Prasad2014} and the Square Kilometer Array (SKA).
\end{itemize}

\begin{figure}
\centering
\includegraphics[width=0.5\textwidth]{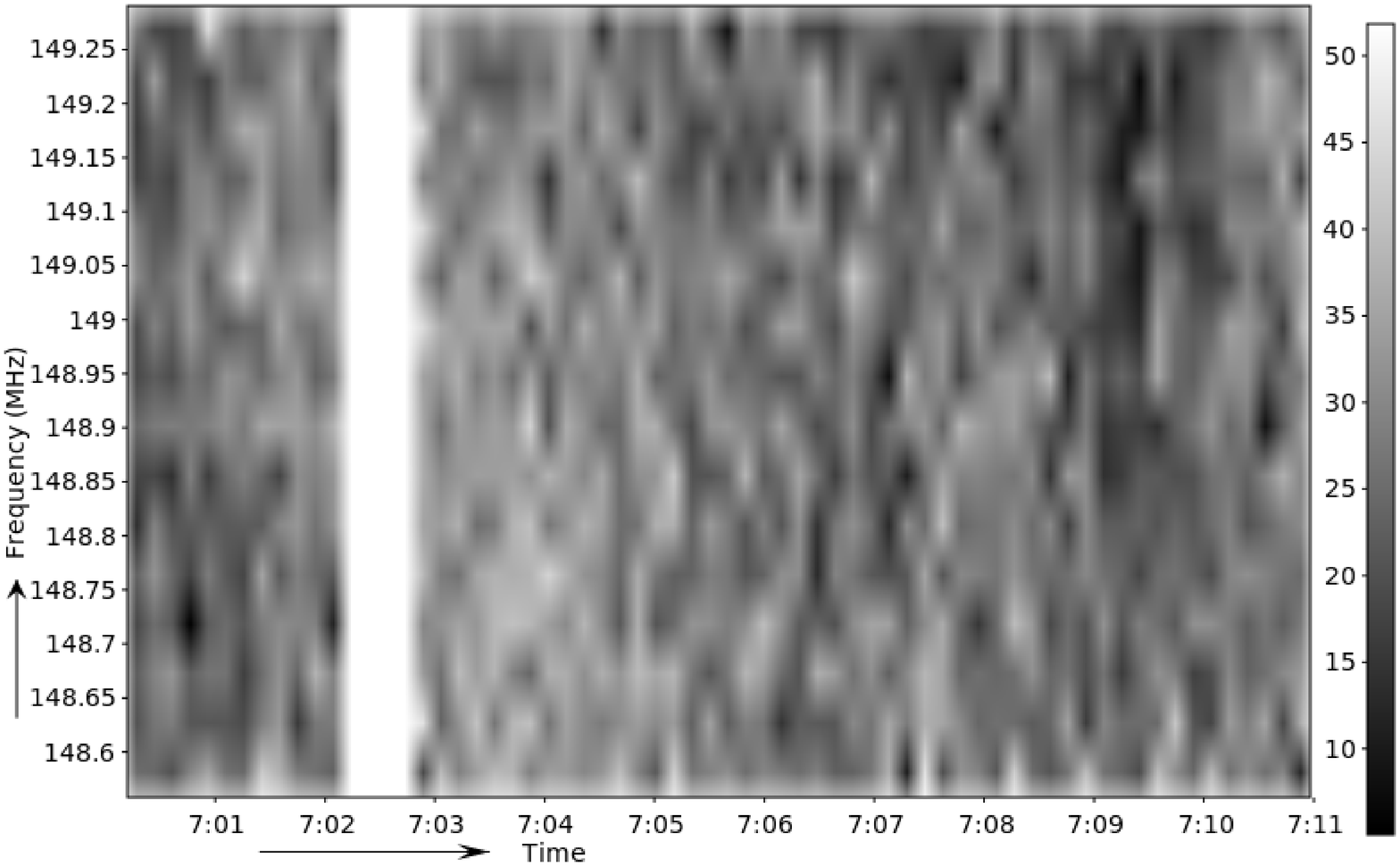}
\centering

\includegraphics[width=0.5\textwidth]{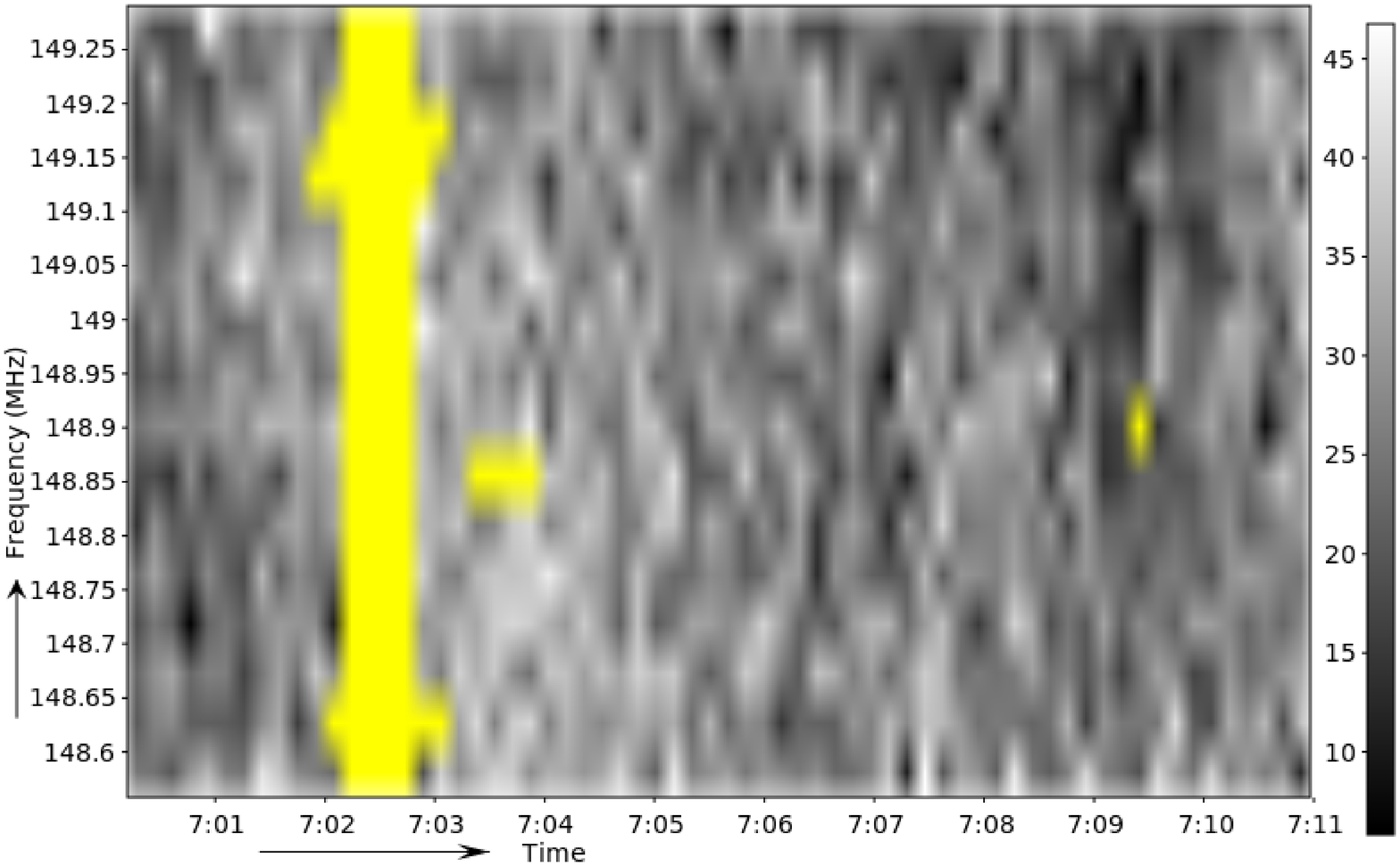}
\centering
\caption{A visual representation of the observed amplitude of a signal in the time-frequency domain on two antennas in the program "rfigui," where the top is an unflagged data set and the bottom has been flagged by AOFlagger (yellow).  Here we have a 30 second step function increase in amplitude from an injected signal approximately a quarter of the way through the data set, which was flagged as RFI.  In this case, however, the flagged signal here is \emph{not} RFI but instead a simulated 30 second transient signal in the data, showing the potential perils of using automatic flagging during transient searches.}
\end{figure}

Overall for routine data collection and subsequent imaging the AOFlagger has proven to be a fast and accurate automatic flagger, particularly in the high-RFI LOFAR radio environment, although it can flag data that is not RFI \citep{Offringathesis}. Normally, however, this is not an issue in routine LOFAR observations because mistakenly flagged data is a small percentage of the entire data sample.  Potential issues however can occur when searching for transient signals that can mimic RFI in their sudden and bright nature-- transient signals are susceptible to automatic flagging algorithms because of their similarities to these unwanted RFI signals and thus may be flagged out before the observer can even look for them.

\section{Transient Simulation Results}

In order to simulate what a transient radio signal would look like in the data, we first took the data from a single measurement set (11 minutes long at 149 MHz, with a bandwidth of 781 kHz).  For these simulations a transient source was injected in the middle of the image at the coordinates of Swift J1644+57 by taking a model point source, transforming it to visibility space, and adding this to the recorded visibilities.  This was done using BlackBoard Selfcal (BBS) software \citep{Loose2008} where the point source was a step function signal of a given strength and duration in the image's sky model.  After this injection, the data were processed as usual through the imaging pipeline described in Section 2, and the resulting image was then analyzed.

This transient's amplitude was varied from 0.5-10 Janskys in half Jansky intervals, and the duration was varied from one second to 11 minutes on a logarithmic scale.  The resulting images for these data were then run through the sourcefinding script PySE where the flux at the location of the injected transient was detected via a forced fit of a point source.  The results of these tests can be seen in Figure 3, where we see that a transient on for the entire measurement's duration would have the specified simulated amplitude, but a transient on for a more brief period of time would have its measured flux decrease as it was averaged out over the total measurement timespan.

\begin{figure}
\centerline{%
\includegraphics[width=0.5\textwidth]{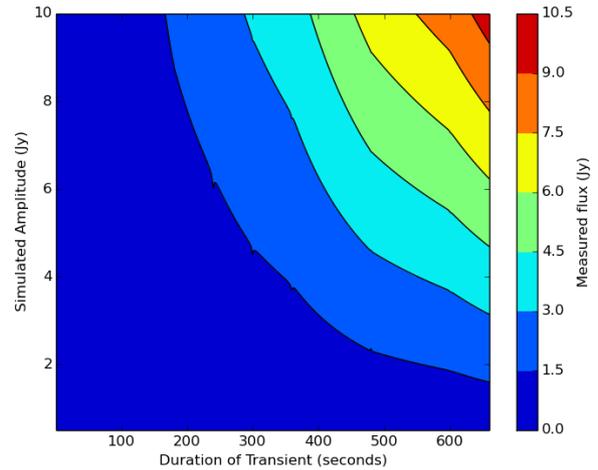}}%
\caption{Observed fluxes for transient signals with a top hat shape of a given amplitude and duration without any secondary automatic flagging over the duration of one measurement set where the maximum length is the 11 minute length of the measurement set.  Here we see that a 10 Jy signal that is on for the entire duration of the image corresponds to a flux level of 10 Jy, but a transient with a shorter $\Delta$t has a correspondingly lower measured flux.}
\end{figure}

After this, the same tests were conducted but with an automatic flagging step added using the standard AOFlagger settings after the transient was injected into the data in order to measure the effects of RFI flagging on the transient signal.  It should also be noted that the data we injected the transient into had already been run through RFI processing during pre-processing when it was first obtained for analysis---  that is, the typical RFI processing for LOFAR imaging.  The tests we describe here involve a secondary RFI flagging step after the transient was simulated and injected into the image but before imaging in order to study the effect of the flagger on this one injected transient signal in particular.  Further, we should note that the data that underwent initial flagging by the LOFAR Observatory was compressed in time and frequency immediately after this stage, but this does not have an effect on this injection and secondary flagging test.

In these tests we found that if the simulated transient was of a longer duration than two minutes there was no statistically significant difference in the observed flux of the transient\textemdash that is, the transient was unaffected by flagging algorithms because its long duration would not trigger the thresholding algorithms.  However, differences could be observed in transients of less than two minutes, and the results of these transients of brief duration can be seen in Figure 4.  If automated flagging is used, very brief transients ($\simeq \Delta t < 10$ s) of a detectable brightness will be flagged out altogether by the automated flagging software once it was bright enough to be detected.  For longer transients (10 s < $\Delta t < 2$~min) the transient is still detectable but at a lower flux, as part of the signal will be flagged.  As outlined earlier, this is relevant for short pulses of coherent radiation that could be emitted due to the jet interaction from the substructure of a TDE, as the shortest duration pulses could be accidentally flagged and never seen by the observer in the first place.

From this, we conclude that radio surveys for brief transient signals relying on automated RFI flagging such as AOFlagger may well be missing the shortest transients.  Further, additional care must be taken for transients of $<2$ min duration after they are identified when it comes to the measured flux of the source because the signal is likely partially flagged during pre-processing.

\begin{figure}
\centering
\includegraphics[width=0.5\textwidth]{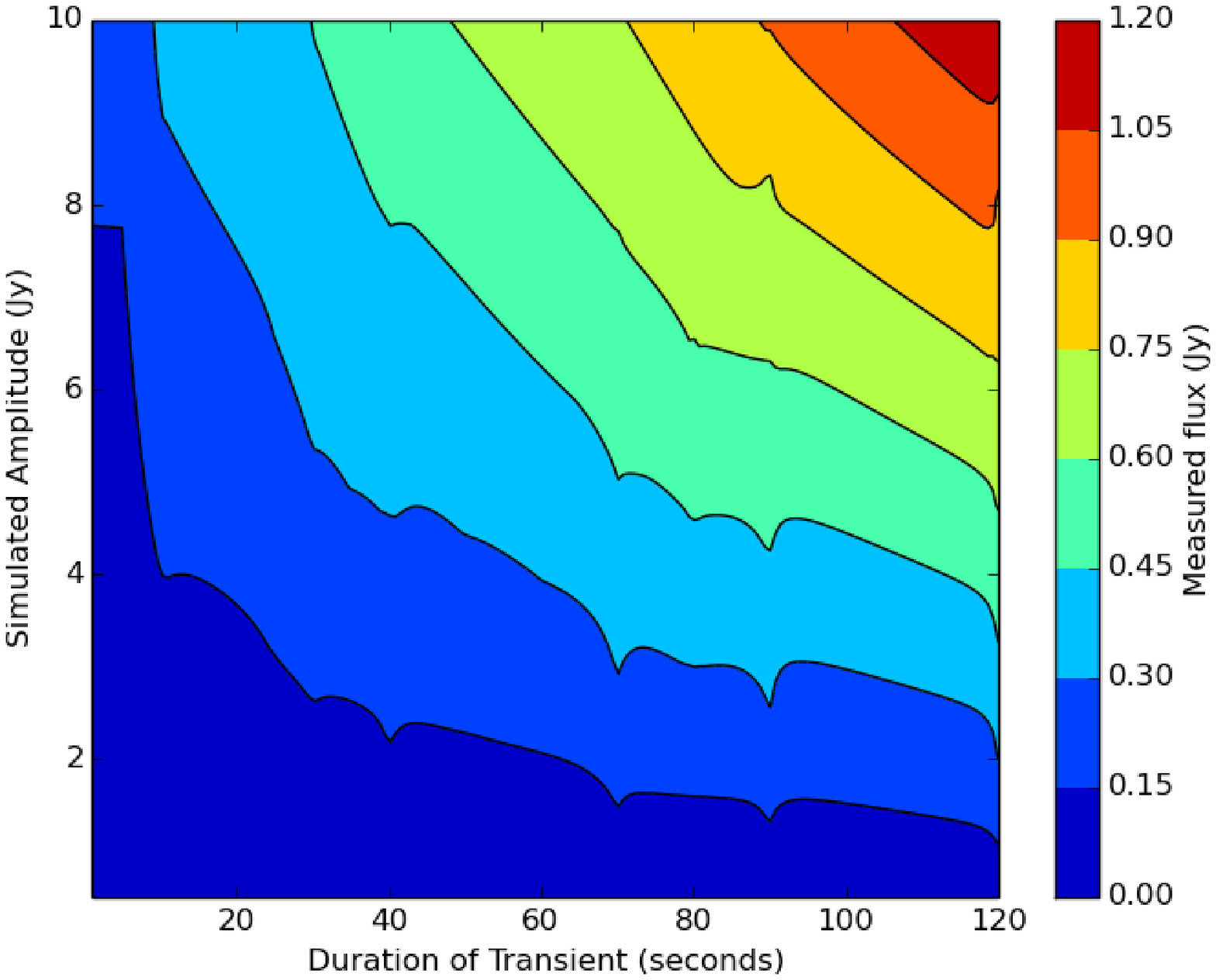}
\centering

\includegraphics[width=0.5\textwidth]{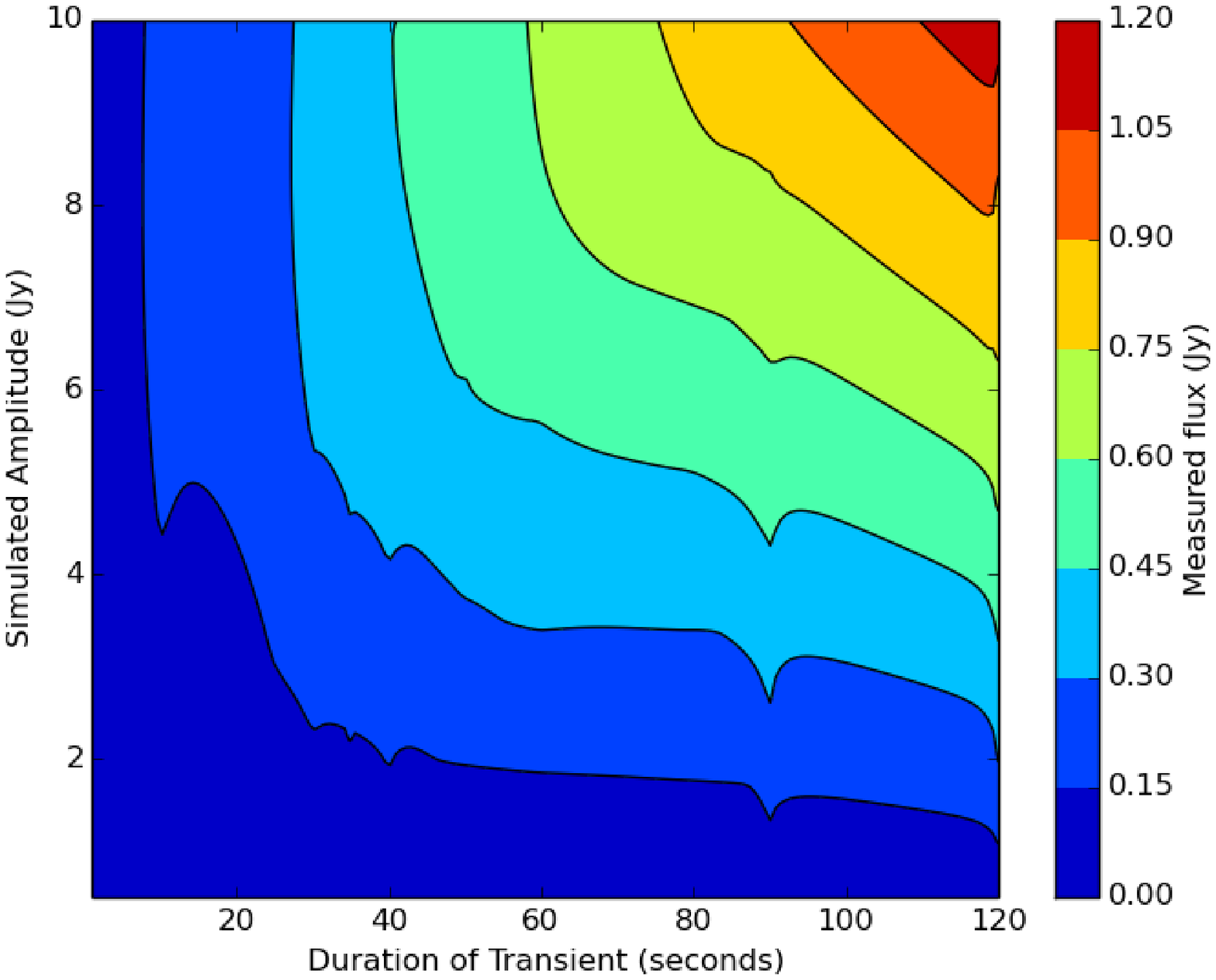}
\centering
\caption{Observed fluxes for transient signals of a given amplitude and duration both with and without automatic flagging over 11 minute integrations.  Here there is an injected transient duration of <2min for both unflagged (top) and flagged (bottom) data.  The random fluctuations in the images here are a sampling effect.}
\end{figure}

In addition to these simulations, we also searched for brief transient sources both within the field of observation and for transients specifically at the TDE location.  The images from the individual Swift J1644+57 measurement sets were run through the Transients Pipeline (Swinbank et al. 2014, in prep) in order to see whether there were any transients detected in the data.  However, no short-duration transients were observed either at the position of the TDE or elsewhere in the field.

\section{Discussion}

Due to its wide field of view and high sensitivity, LOFAR is a prime candidate to survey both for TDEs and brief transient signals.  Currently in the routine LOFAR transient monitoring (Fender et al., in prep) there is an average sensitivity limit of roughly 100 mJy at a 5 sigma confidence level at 149 MHz for an 11 minute snapshot with a maximum baseline of 12 km, and there is a total survey area of 1,500 square degrees.  As outlined in \citet{Berger2012}, the plateau in emission from Swift J1644+57 occurred $\sim$600 days after its initial outburst.  Assuming a consistent peak emission synchrotron spectrum as outlined in \citet{Zauderer2013}, where the peak frequency follows the relationship $v_{p} \propto t^{-1.3}$ and peak flux $F(v_{p}) \propto{} t^{-0.8}$, a TDE at 150 MHz would occur $\sim$10 years after the initial event and would have an observed flux of 0.2 mJy.  This would require the TDE to be a fairly energetic event to be detectable, but with the added advantage that events similar to Swift J1644+57 could be viewed years later during the non-relativistic phase without an initial triggering event such as that of Swift J1644+57.  It should also be noted that in a dedicated observation of 12 hours, a LOFAR observation with 6 kilometer baselines can reach down to the mJy level, and using the full high band (~115-190 MHz) sub-mJy noise levels can be achieved.

Further, due to the null detection of transients within our data, we can constrain the snapshot rate of transient events during our observation by using the method described in \cite{Carbone2014} which sets an upper limit on the number of transient events.  First, assuming a Poisson distribution of sources in order to get the 95\% upper limit on the snapshot rate $\rho$ within our data, we can use the relationship
\begin{equation}
P(0) = e^{-\rho\Omega_{\rm{tot}}} = 1-C
\end{equation}
where P(0) is when no transients were detected, C is our confidence level of 0.95, and $\Omega_{\rm{tot}}$ is the field of view in square degrees at the observing frequency (11.35 deg$^2$) multiplied by the number of snapshots taken of the field over the course of our observations.  From this, we get a limit of $\rho < 2.2 \times 10^{-2}$ deg$^{-2}$.

Further, as explained in detail by \cite{Carbone2014}, we can assume the number density of transient sources has a power-law distribution in flux as follows:

\begin{equation}
N(S > \hat{S} )= N_{*} \frac{\hat{S}}{S_{*}}^{-\gamma}
\end{equation}

where $S_{*}$ is an arbitrary value of the flux at which the normalization is given (which we will take here to be 0.5 Jy) and $\gamma$ is the exponent upon which the power law distribution of sources is dependent. From this, we can find the upper limit for the number of transient sources as a function of our sensitivity.  This relationship for the normalization $N_{*}$ is

\begin{equation}
N_{*} < -\frac{\rm{ln}{(0.05)}}{\Omega} \left(\frac{S_{*}}{D}\right)^{-\gamma} \frac{1}{\sum{\sigma_{i}^{-\gamma}}}
\end{equation}

where $\Omega$ is now the field of view of one snapshot, D is our signal-to-noise threshold, and $\sigma_{i}$ is the noise in a single image.  The limits of our transient survey for the snapshot surface density in terms of the exponent of the assumed flux distribution can be seen in Table 2.  These limits fall above the upper limits of recent radio transient surveys \citep{Murphy2014};\citep{Frail2012}.

\begin{table}
\centering
\begin{tabular}{lccc}
\hline
$\gamma$&Upper limit of snapshot surface density (deg$^{-2}$)\\
\hline
0.0 & 0.264 \\
0.5 & 0.227 \\ 
1.0 & 0.195 \\ 
1.5 & 0.160 \\ 
2.0 & 0.144 \\ 
    \end{tabular}
\caption{\label{t2}The snapshot surface density for different values of the exponent of the assumed flux distribution of transient sources.}
\end{table}

There is also the question of how often we can statistically expect to see short duration transients from a specific location in the sky.  Brief bursts, such as those described by \cite{Usov2000}, are only visible at lower frequencies because the spectrum, with shape $F_{v} \propto \nu^{-1.6}$, is estimated to peak in this regime.  Thus such flashes could not be easily monitored in previous monitoring campaigns of Swift J1644+57 because those campaigns were in the higher frequency regime.

If one such flash with a duration $t_{\rm{signal}}$, occurs during an observation with duration $t_{\rm{obs}}$, it would be averaged out over the course of the observation as follows:

\begin{equation}
F_{\rm{signal}} = \frac{t_{\rm{signal}}}{t_{\rm{obs}}} F_{\rm{o}}
\end{equation}  

 where $F_{\rm{signal}}$ is the observed signal strength measured from the source over the course of the entire data set in analysis, and \emph{$F_{\rm{o}}$} refers to the original strength of the signal.  When the flux of the source was measured using a forced fit in PySE for each of the images from individual data sets, we observed an average of the flux over all the individual images as $-0.015 \pm 0.405$ Jy (in comparison, our previous measurement described in Section 2 was for one image where all the data was concatenated together into one deeper image).  A scatter plot of the fluxes measured over the images can be seen in Figure 5.

\begin{figure}
\includegraphics[width=0.5\textwidth]{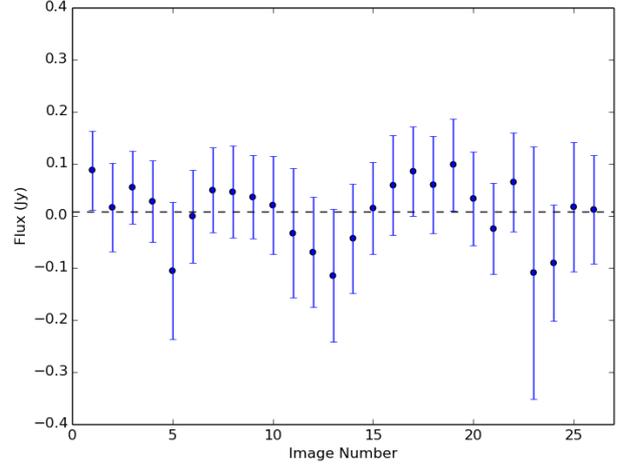}
\centering
\caption{A plot of the flux of the source in Janskys measured in each individual image.  The dotted line is at the average for the integrated image at 0.0083 Jy, shown here for reference.}
\end{figure}

From the simulation results discussed in the previous section, we can then use this $F_{\rm{signal}}$ value to constrain some other parameters in this equation.  Assuming the simplest case of one such transient in an image, we can apply a cutoff of all simulated transients exceeding this value based on their signal strength and duration, as seen in Figure 6.  Using this model we can see that bright, brief transients on the order of a few seconds up to a flux of a few Janskys could have gone undetected, but we can exclude the possibility of longer and brighter transients having occured in the data.  For the faintest transient signals (on the order of less than half a Jansky), such transients could not have exceeded 2 minutes duration during our monitoring of the Swift J1644+57 field.

\begin{figure}
\includegraphics[width=0.5\textwidth]{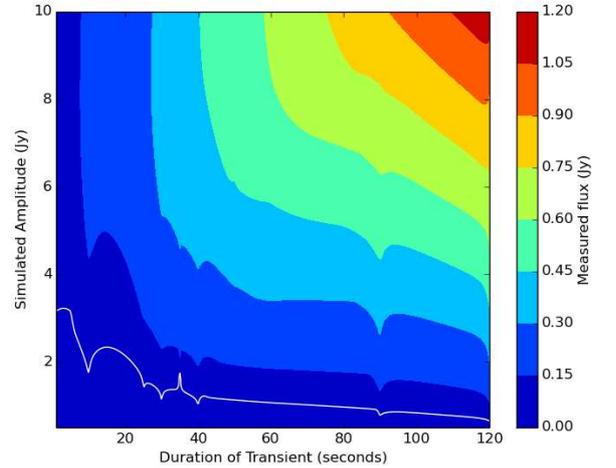}
\centering
\caption{Observed fluxes for simulated transient signals of a given amplitude and duration in a specific location, along with the cutoff line (in white) above which no transient signals were observed during our observations.}
\end{figure}

As outlined in the previous section, this estimate is also hampered by the effects of routine automatic RFI flagging during the pre-processing of our data: signals below a 10 second duration will be flagged altogether and those of a 10 second --- 2 minute duration are partially flagged in the current data, further diminishing the observation of such transients.  This also hampers conclusions on scenarios where multiple bursts of radiation occur during the course of an observation, such as if multiple flashes of a one second observation were observed over the course of several minutes.  Even if $t_{\rm{signal}}$ for the transient is longer in such a scenario, the fact that it is spread out over different parts of the measurement mean the brief flashes are likely to be flagged as RFI.  Further, because such flashes are so difficult to distinguish from real RFI, it is very difficult to study them in great detail.

Finally, there are other factors to investigate when it comes to transient signals and RFI flagging that are outside of the scope of this paper.  First, transients can have a spectral width and evolution that the automated flagger does not take into account during its data processing, and some transients may have a signal dispersed in frequency.  This could make investigation worthwhile into different uv cuts in the data in order to see how the RFI rejection works in relation to transients.  Second, for transient searches looking for signals of less than 10 seconds duration, it may be necessary to image the unflagged data in order to see whether there are any transient signals of this duration detected that would be mistaken for RFI otherwise.  It is worth considering whether an RFI flagger could in fact aid transient detection of these sources by acting as a transient finder.  A transient would appear in the flagger as flagged data short in the time domain but broad in the frequency domain, but originates from one point in the sky and converge to a point source upon imaging.  As such, flagged data that matches these criteria when imaged may yield transient sources.  RFI, on the other hand, would most likely have different origins, meaning it would result in a noise-filled image.

\section{Conclusions}

We analyzed low frequency radio data of TDE event Swift J1644+57, and found a non-detection with a peak flux density at this position of 24.7 $\pm$ 8.9 mJy.  We also considered the possibility of short duration transients within the data set, both within the general field covered by the data and brief transients that could originate from the TDE event location itself.  Our upper limit on the transient rate of the snapshot surface density in this field at sensitivities < 0.5 Jy is $\rho < 2.2\times10^{-2}$ deg$^{-2}$, and we conclude that we did not observe any brief transient signals originating specifically from the Swift J1644+57 source itself to a level of $-0.015 \pm 0.405$ Jy.  

However, we have also demonstrated that for transients of less than 2 minutes duration automatic RFI flagging will affect these signals by flagging out either part of or the entire transient signal.  This means that transient signals of greater than 2 minutes duration will survive the flagging process, and we have a good chance of observing them, albeit the original data would need to be reexamined in order to determine whether part of the signal was mistakenly flagged (thus diminishing the transient's observed flux).  Additional investigations taking into account the spectral shape of  transient signals, and whether there are any transients in already flagged RFI data, could shed more light on this topic.

\section*{Aknowledgements}

We acknowledge support from the European Research Council via the Advanced Investigator Grant no. 24729.  This work is also supported in part by European Research Council Advanced Grant 267697.  LOFAR, the Low Frequency Array designed and constructed by ASTRON, has facilities in several countries, that are owned by various parties (each with their own funding sources), and that are collectively operated by the International LOFAR Telescope (ILT) foundation under a joint scientific policy.  We would like to thank the LOFAR Observatory staff for their assistance in obtaining and the handling of this large data set.  S.C. acknowledges funding support from the UnivEarthS Labex program of Sorbonne Paris Cit\'e (ANR-10-LABX-0023 and ANR-11-IDEX-0005-02).

\end{document}